\documentclass[12pt,a4paper]{article}
\usepackage{mitpress}

\newdimen\dummy
\dummy=\oddsidemargin
\input{tcilatex}

\begin{document}

\title{{\large Infrared study of spin-Peierls compound }$\alpha ${\large 
\'{}%
-NaV}$_{2}${\large O}$_{5}\thanks{%
Contributed paper for the SCES98 (15-18 July 1998, Paris, France). To be
published in Physica B.}$}
\author{{\small D. Smirnov }$^{a}\thanks{%
e-mail: smirnov@insa-tlse.fr}${\small , J. Leotin }$^{a}${\small , P. Millet 
}$^{b}${\small , J. Jegoudez }$^{c}${\small , A. Revcolevschi }$^{c}\medskip 
$ \\
$^{a}${\scriptsize \ Laboratoire de Physique de la Mati\`{e}re
Condens\'{e}e, INSA-SNCMP, avenue de Rangueil, 31077 Toulouse, France}\\
$^{b}${\scriptsize \ Centre d'Elaboration de Mat\'{e}riaux et d'Etudes
Structurales, 29 rue Jeanne Marvig, 31055 Toulouse, France}\\
$^{c}${\scriptsize \ Laboratoire de Chimie des Solides, Universit\'{e} de
Paris-sud, B\^{a}t. 414, F-91405 Orsay, France}}
\maketitle

\begin{abstract}
{\scriptsize Infrared reflectance of }$\alpha 
{\acute{}}%
${\scriptsize -NaV}$_{2}${\scriptsize O}$_{5}${\scriptsize \ single crystals
in the frequency range from 50 cm}$^{-1}${\scriptsize \ to 10000 cm}$^{-1}$%
{\scriptsize \ was studied for }$a${\scriptsize , }$b${\scriptsize \ and }$c$%
{\scriptsize -polarisations. In addition to phonon modes identification, for
the }$a${\scriptsize -polarised spectrum a broad continuum absorption in the
range of 1D magnetic excitation energies was found. The strong near-IR
absorption band at 0.8 eV shows a strong anisotropy with vanishing intensity
in }$c${\scriptsize -polarisation. Activation of new phonons due to the
lattice dimerisation were detected below 35K as well as pretransitional
structural fluctuations up to 65K.}
\end{abstract}


{\small \medskip }

{\small Recently, a spin-Peierls (SP) transition was found in inorganic
compounds CuGeO}$_{3}${\small \ and }$\alpha 
{\acute{}}%
${\small -NaV}$_{2}${\small O}$_{5}${\small \ \cite{Hase,Isobe+Ueda}. The
evidence of structural phase transition in }$\alpha 
{\acute{}}%
${\small -NaV}$_{2}${\small O}$_{5}${\small \ was obtained , first, by
observation of additional superlattice X-ray reflections \cite{Fujii}, and
then by measurements of new phonon lines in infrared (IR) \cite
{Popova,Smirnov} and Raman \cite{Weiden} spectra below T}$_{SP}\approx $%
{\small 35K. The 1D magnetic properties of }$\alpha 
{\acute{}}%
${\small -NaV}$_{2}${\small O}$_{5}${\small \ were explained on the basis of
the non-centrosymmetric space group }$P_{2_{1}mn}${\small \ as initially
determined \cite{Carpy+Galy}.}

{\small Room temperature IR reflectance was measured in the frequency range
between 50 cm}$^{-1}${\small \ and 10000 cm}$^{-1}${\small \ on two }$\alpha 
{\acute{}}%
${\small -NaV}$_{2}${\small O}$_{5}${\small \ single crystals grown
according to the procedure described in Ref. \cite{Isobe'growth}. For
transmission measurements }$\alpha 
{\acute{}}%
${\small -NaV}$_{2}${\small O}$_{5}${\small \ powder was ground in KBr and
pressed in pellets. The IR measurements were done with a Bruker IFS 113V
Fourier spectrometer with a resolution ranging from 1 cm}$^{-1}${\small \ in
the far-IR region to 5 cm}$^{-1}${\small \ in the near-IR region. The low
temperature spectra were measured with a continuous He flow cryostat with
the absolute accuracy of temperature control about 0.1 K. Reference spectra
(reflectance of the freshly evaporated gold film or direct transmittance)
were measured after each sample spectrum measurement.}

{\small The 300K polarised reflectivity R(}$\nu ${\small ) spectra of }$%
\alpha 
{\acute{}}%
${\small -NaV}$_{2}${\small O}$_{5}${\small \ are shown in Fig.1. From the
factor group analysis one should expect 37 IR active phonon modes for the
non-centrosymmetric structure: }$\Gamma (P_{2_{1}mn})_{\text{IR}%
}=15A_{1}(E\parallel a)+7B_{1}(E\parallel b)+15B_{2}(E\parallel c)${\small ,
or 18 modes for the recently proposed \cite{Smolinski} centrosymmetric one: }%
$\Gamma (P_{mmn})_{\text{IR}}=7B_{1u}(E\parallel c)+4B_{2u}(E\parallel
b)+7B_{3u}(E\parallel a)${\small . We observed 6 }$a${\small -polarised, 4 }$%
b${\small -polarised and 5 }$c${\small -polarised phonons. This information
does not allow to choose between two space groups since some phonons may
have too small oscillator strengths to be detected.}

{\small In order to quantitatively describe R(}$\nu ${\small ) spectra, we
applied the classical approach based on a harmonic oscillator model assuming
that all transitions (either electronic or phonon) provide a Lorenzian
contribution to the dielectric function: } 
\[
\varepsilon (\nu )=\varepsilon _{\infty }+\stackunder{j}{\sum }\frac{\nu
_{S,j}^{2}}{\nu _{j}^{2}-\nu ^{2}-i\cdot \gamma _{j}\cdot \nu } 
\]
{\small where }$\varepsilon _{\infty }${\small \ is the high frequency
dielectric constant; }$\nu _{j}${\small , }$\nu _{S,j}${\small \ , }$\gamma
_{j}${\small \ are the frequency, the oscillator strength and the damping of
the j-th oscillator (for the phonons }$\nu _{j}${\small =}$\nu _{TO}${\small %
). Calculated reflectivity spectrum is fitted to the experimental one
starting from the }$\nu _{j}${\small , }$\nu _{S,j}${\small \ , }$\gamma
_{j} ${\small \ values determined using Kramers-Kronig analysis. In the
region of phonon frequencies (}$\nu ${\small \ \TEXTsymbol{<} 1000 cm}$^{-1}$%
{\small ), the R(}$\nu ${\small ) spectra have a typical form of phonon
bands in an insulating material only for b- and c- polarisation. In order to
describe the R(}$\nu ${\small ) spectrum for a-polarisation, it is necessary
to introduce in the dielectric function two wide Lorenzian ''continuum''
bands centred at approximately 280 cm}$^{-1}${\small \ and 1100 cm}$^{-1}$%
{\small . Let us note that these energies (}$\approx ${\small 400 K and
1580K, respectively) fall in the energy range of magnetic excitations in a
Heisenberg S=1/2 spin chain (the nearest-neighbour intrachain constant in }$%
\alpha 
{\acute{}}%
${\small -NaV}$_{2}${\small O}$_{5}${\small \ is }$J\approx ${\small 440}$%
\div ${\small 560 K \cite{Weiden,Isobe+Ueda} ). This ''continuum'' might
result from a direct two-magnon absorption mechanism recently suggested in
Ref. \cite{Damasc}. It could also involve a phonon assisted two-magnon
scattering process \cite{Lore+Saw}. On the other hand, the near-IR strong
optical transition found for E}$\parallel ${\small a was explained recently
as a charge transfer (CT) absorption mechanism between two on-rung V sites 
\cite{Damasc}. Following this approach, one can calculate the energy
difference between V sites on the same rung (a-direction ) }$\Delta \approx $%
{\small 0.6 eV and the hopping constant }$t_{\perp }\approx ${\small 0.33
eV. Then, the estimated valence state of the vanadium atoms is 4.2 and 4.8.}

{\small All phonon modes with }$\nu _{TO}${\small \TEXTsymbol{>}200 cm}$%
^{-1} ${\small \ except the weak E}$\parallel ${\small c mode at 583 cm}$%
^{-1}${\small \ were also observed in transmission spectra of }$\alpha 
{\acute{}}%
${\small -NaV}$_{2}${\small O}$_{5}${\small \ powder dispersed in KBr
pellets . Below }$T_{SP}${\small \ we detected 6 new lines at 363, 411, 716,
969, 1275, 1399 cm}$^{-1}${\small . The pellets technique is especially
useful to analyse phonon modes with different oscillator strength. We use
this method to complete the study of the evolution with temperature of the
strongest new phonon line at }$\approx ${\small 718 cm}$^{-1}${\small \ \cite
{Smirnov}. Detailed temperature dependence of the above line shown in Fig. 2
gives a clear evidence of precursor effects up to at least 64K. The
observation of pretransitional spectral signatures at }$T${\small 
\TEXTsymbol{>}}$T_{SP}${\small \ linked to the new phonon in the SP state
indicates the existence of local structural fluctuations likely driven by 1D
magnetic fluctuations via strong magnetoelastic coupling.}

{\small In conclusion, first IR measurements of }$\alpha 
{\acute{}}%
${\small -NaV}$_{2}${\small O}$_{5}${\small \ along all three }$a${\small , }%
$b${\small \ and }$c${\small -polarisations are reported. An anomalous
behaviour along }$a${\small \ -polarisation is evidenced by both a low
frequency continuum involving spin excitations and near-IR charge transfer
band. Pretransitional structural fluctuations are observed in a wide
temperature range above }$T_{SP}${\small .}

{\small We acknowledge to D. van der Marel for useful discussions.}

{\small \bigskip }

{\small Table 1. Oscillator parameters (in cm}$^{-1}${\small ) used to fit
the IR reflectivity spectra of }$\alpha 
{\acute{}}%
${\small -NaV}$_{2}${\small O}$_{5}${\small \ .}

\begin{tabular}{|p{2cm}|p{2cm}|p{2cm}|p{2cm}|p{2cm}|p{2cm}|}
\hline
\multicolumn{6}{|p{12cm}|}{\textbf{E}$\parallel $\textbf{a} ($\varepsilon
_{\infty }$ = 2.6)} \\ \hline
$\nu _{TO}$ & 940 & 522 & 252 & 146.7 & (*) \\ \hline
$\nu _{S,TO}$ & 42 & 853 & 87 & 62 &  \\ \hline
$\gamma _{TO}$ & 8.2 & 28.4 & 9.6 & 8.8 &  \\ \hline
\multicolumn{6}{|p{12cm}|}{{\small ``Continuum`` contribution: 1). }$\nu $%
{\small \ =278, }$\nu _{S}${\small \ = 472, }$\gamma ${\small \ = 209; 2). }$%
\nu ${\small \ =1109, }$\nu _{S}${\small \ =647, }$\gamma ${\small \ =522.
CT contribution: 1). }$\nu ${\small =3970, }$\nu _{S}${\small \ =4120, }$%
\gamma ${\small =2260; 2). }$\nu ${\small =6540, }$\nu _{S}${\small \
=11970, }$\gamma ${\small =2550}} \\ \hline
\end{tabular}

{\small \smallskip }

\begin{tabular}{|p{2cm}|p{2cm}|p{2cm}|p{2cm}|p{2cm}|p{2cm}|}
\hline
\multicolumn{6}{|p{12cm}|}{\textbf{E}$\parallel $\textbf{b} ($\varepsilon
_{\infty }$ = 1.5)} \\ \hline
$\nu _{TO}$ & 580 & 366 & 230 & 176 &  \\ \hline
$\nu _{S,TO}$ & 1046 & 291 & 32 & 107 &  \\ \hline
$\gamma _{TO}$ & 14.9 & 12.1 & 10.1 & 6.8 &  \\ \hline
\multicolumn{6}{|p{12cm}|}{CT contribution: $\nu $=7280, $\nu _{S}$ =12780, $%
\gamma $=7160} \\ \hline
\end{tabular}

{\small \smallskip }

\begin{tabular}{|p{2cm}|p{2cm}|p{2cm}|p{2cm}|p{2cm}|p{2cm}|}
\hline
\multicolumn{6}{|p{12cm}|}{\textbf{E}$\parallel $\textbf{c} ($\varepsilon
_{\infty }$= 3.0)} \\ \hline
$\nu _{TO}$ & 954 & 583 & 468 & 182 & 164 \\ \hline
$\nu _{S,TO}$ & 589 & 90 & 219 & 207 & 55.6 \\ \hline
$\gamma _{TO}$ & 3.7 & 58 & 36.9 & 10.5 & 3.2 \\ \hline
\end{tabular}

{\small * - lines at 740 cm}$^{-1}${\small \ and 90 cm}$^{-1}${\small \ are
observed at T}$<${\small 100K}

{\small \bigskip }

{\small Figure captions.\medskip }

{\small Fig. 1. 300K polarised reflectivity spectra of }$\alpha 
{\acute{}}%
${\small -NaV}$_{2}${\small O}$_{5}${\small \ single crystal. The solid
lines represent the calculated spectra. Not all the experimental points are
shown for the sake of clarity.\medskip }

{\small Fig. 2. Temperature dependence of the (716-718) cm}$^{-1}${\small \
line. Open circles and squares indicate the integral intensity of the
transmittance (716 cm}$^{-1}${\small ) and reflectivity (718 cm}$^{-1}$%
{\small ) line, respectively. The solid triangles show the intensity of
Lorenzian component deduced from reflectivity data as described in Ref. \cite
{Smirnov}.}

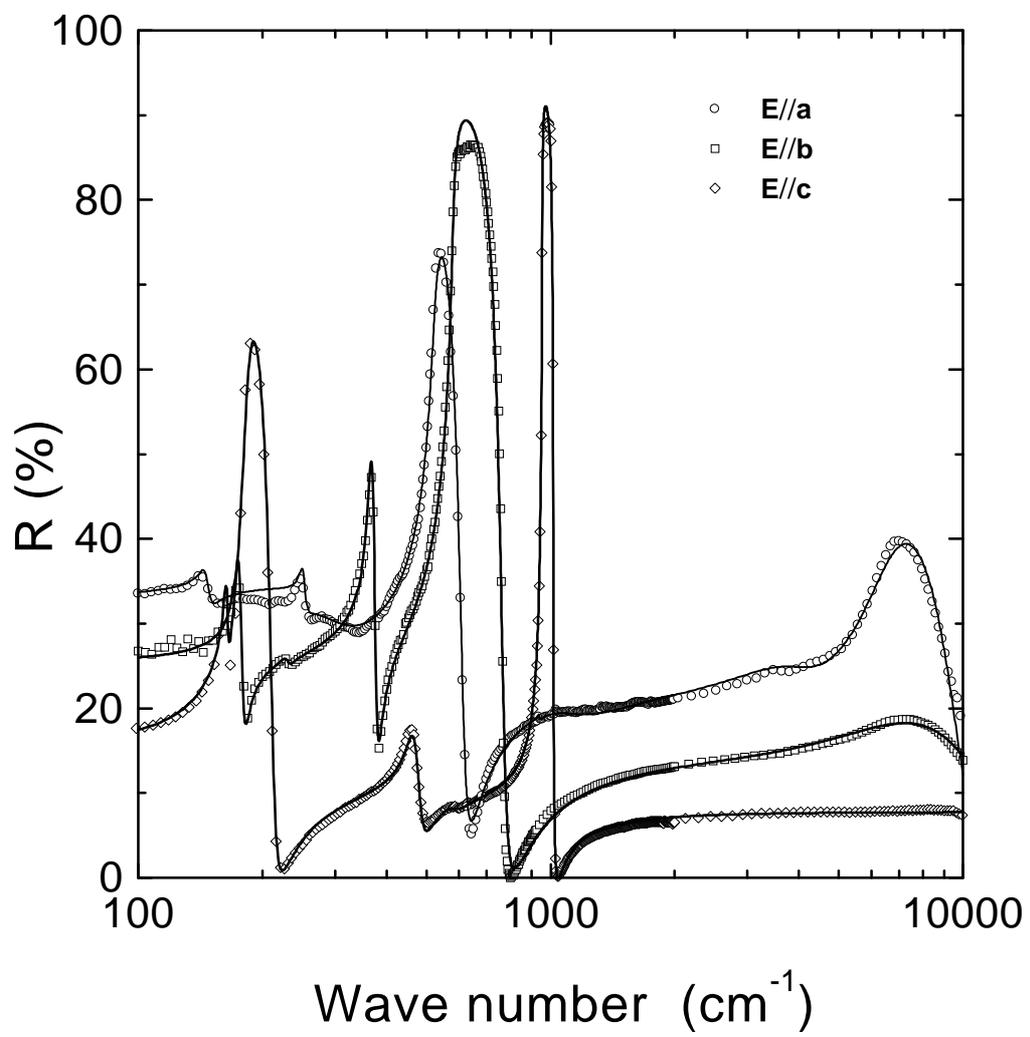
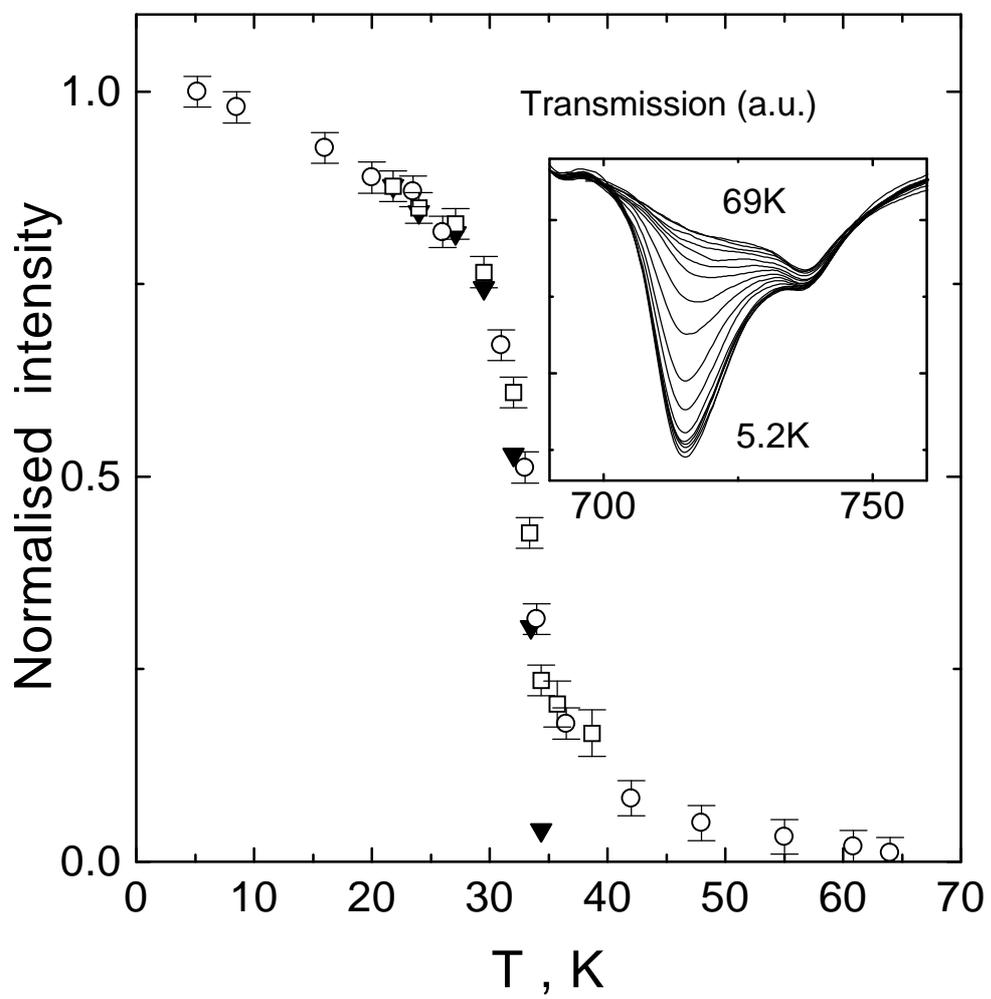
\end{document}